\documentclass[fleqn,usenatbib]{mnras}

\usepackage{newtxtext,newtxmath}
\usepackage[T1]{fontenc}
\usepackage{ae,aecompl}

\usepackage{graphicx}	% Including figure files
\usepackage{amsmath}	% Advanced maths commands
\usepackage{amssymb}	% Extra maths symbols

\usepackage{times}
\newcommand{\vect}[1]{\boldsymbol{#1}}
\newcommand{\diff}{\text{d}}
\date{}

\pubyear{2017}

\title[Dynamical tides in NSs with hyperon cores]{Dynamical tides in coalescing superfluid neutron star binaries with hyperon cores and their detectability with third generation gravitational-wave detectors}

\author[H. Yu and N. N. Weinberg]{Hang Yu$^{1, 2}$\thanks{E-mail: hyu45@mit.edu} and Nevin N. Weinberg$^{1}$\\
$^{1}$Department of Physics, and Kavli Institute for Astrophysics and Space Research, Massachusetts Institute of Technology, \\Cambridge, MA 02139, USA\\
$^{2}$LIGO Laboratory, Massachusetts Institute of Technology, Cambridge, MA 02139, USA}

\begin{document}
\label{firstpage}
\pagerange{\pageref{firstpage}--\pageref{lastpage}}
\maketitle

% Abstract of the paper
\begin{abstract}
The dynamical tide in a coalescing neutron star binary induces phase shifts in the gravitational waveform as the orbit sweeps through resonances with individual g-modes.  
Unlike the phase shift due to the equilibrium tide, the phase shifts due to the dynamical tide are sensitive to the stratification, composition, and superfluid state of the core.  We extend our previous study of the dynamical tide in superfluid neutron stars by allowing for hyperons in the core.  Hyperon gradients give rise to a new type of composition g-mode. Compared to g-modes due to muon-to-electron gradients, those due to hyperon gradients  are concentrated much deeper in the core and therefore probe higher density regions.  
We find that the phase shifts due to resonantly excited hyperonic modes are $\sim$$10^{-3}\textrm{ rad}$, an order of magnitude smaller than those due to muonic modes.
We show that by stacking events, third generation gravitational-wave detectors should be able to detect the phase shifts due to muonic modes. Those due to hyperonic modes will, however, be difficult to detect due to their smaller magnitude.  
%unless there is an event within the Local Group, or there are multiple third generation detectors and the merger rates are on the high end of the predicted range.
\end{abstract}

% Select between one and six entries from the list of approved keywords.
% Don't make up new ones.

%%%%%%%%%%%%%%%%%%%%%%%%%%%%%%%%%%%%%%%%%%%%%%%%%%

%%%%%%%%%%%%%%%%% BODY OF PAPER %%%%%%%%%%%%%%%%%%
\section{Introduction}

Tides in coalescing neutron star (NS) binaries modify the rate of inspiral and generate phase shifts in the gravitational wave (GW) signal that encode information about the the NS interior.   The tide is often decomposed into an equilibrium tide and a dynamical tide, where the former represents the fluid's quasi-static response and the latter represents its resonant response (e.g., in the form of resonantly excited g-modes).  The GW phase shift due to the equilibrium tide, which should be detectable with Advanced LIGO \citep{Aasi:15} by stacking multiple merger events, can constrain the NS tidal deformability and therefore the supranuclear equation of state \citep{Read:09, Hinderer:10, Damour:12, DelPozzo:13, Lackey:15, Agathos:15}.  However, the equilibrium tide can only indirectly constrain the interior stratification (i.e., the composition profile; \citealt{Chatziioannou:15}) and is insensitive to superfluid effects (see, e.g., \citealt{Penner:11}).   By contrast, the dynamical tide is directly sensitive to both the stratification  \citep{Shibata:94, Lai:94,  Reisenegger:94, Kokkotas:95, Ho:99, Hinderer:16, Steinhoff:16} and superfluid effects \citep{Yu:17}.  GW phase shifts due to the dynamical tide can therefore provide a unique probe of the NS interior, similar to asteroseismology observations which are now providing detailed constraints on the physics of the interiors of white dwarfs, solar-type stars, and red giants  \citep{Winget:08, Chaplin:13}.

Previous studies of the dynamical tide in binary NSs focused on `canonical' $1.4 M_\odot$ NSs and assumed that the core does not contain exotic hadronic matter, such as hyperons. Although it  is energetically favorable for nuclear matter to transition to hyperonic matter at high densities \citep{Ambartsumyan:60}, the discovery of $2 M_\odot$ NSs \citep{Demorest:10, Antoniadis:13} ruled out many hyperonic models since they tend to have softer equations of state.   However, the degree of softening is uncertain   \citep{Lonardoni:15} and in the last few years many new hyperonic models compatible with the observations of $2M_\odot$ NSs have been proposed (e.g., \citealt{Bednarek:12, Weissenborn:12, Gusakov:14, Tolos:16}). 

The dynamical tide in hyperonic models is modified by the hyperon composition gradient, which provides a new source of buoyancy that can support g-mode oscillations much deeper within the NS core than the leptonic composition gradient \citep{Dommes:16}. GW phase shifts induced by the excitation of hyperonic g-modes therefore probe the innermost core, where the density is a few times the nuclear saturation density.
 
The total phase shift accumulated over the inspiral due to the equilibrium tide is $\sim 1 \textrm{ rad}$ while that due to the dynamical tide is only  $\lesssim 10^{-2}\textrm{ rad}$.   However, their detectability is not as different as these numbers might suggest.  In part, this is because the dynamical tide phase shift accumulates at lower GW frequencies, where ground-based detectors are more sensitive. There is also more time before the merger to build up the signal-to-noise ratio (SNR) and compare the waveform signal before and after resonance. In addition, because the dynamical tide causes small but sudden increases in GW frequency at mode resonances, it has a unique signature that cannot be easily mimicked by varying other parameters of the binary (such as the masses).

In this paper we extend our previous study of dynamical tides in superfluid NSs \citep{Yu:17} in order to account for the possible presence of hyperons in the core.  We also evaluate the detectability of the phase shifts induced by the dynamical tide with second and third generation GW detectors. We begin in Section \ref{sec:bgModel} by describing our background hyperonic NS model and in Section \ref{sec:eigenmodes} we solve for its g-modes.  In Section \ref{sec:tides}, we consider the resonant tidal driving of the g-modes  and calculate the resulting GW phase shift.  In Section \ref{sec:detectability}, we evaluate the prospects for detecting the phase shifts with current and future GW detectors.

\section{SUPERFLUID MODELS with hyperons}
\label{sec:bgModel}

We construct our background superfluid models using an approach similar to that of \citeauthor{Yu:17} (2017; hereafter YW17), but with some key differences described below; we refer the reader to YW17 and references therein for further details, particularly as pertains to our treatment of the thermodynamics.  Briefly, we assume that the background star is non-rotating, cold (zero temperature), in chemical equilibrium, and that all charge densities are balanced.  We treat the neutrons in the core as superfluid and all other particle species  (protons, $\Lambda$-hyperons, electrons, and muons) as normal fluid matter.\footnote{The conditions under which $\Lambda$-hyperons become superfluid in NS cores are uncertain \citep{Takatsuka:06, Wang:10}.  We assume that they are normal fluid in this study, similar to the g-mode calculations in \citet{Dommes:16}.} 

The two main differences between the present approach and that of YW17 are that here: (1) we use the GM1'B  equation of state \citep{Gusakov:14} rather than SLy4\citep{Stone:03}, and (2) we solve the general relativistic (GR) Tolman-Oppenheimer-Volkhov (TOV) equations of stellar structure rather than the Newtonian equations. We use GM1'B because it allows for the existence of hyperons in the inner core, is consistent with the existence of $2 M_\odot$ NSs,  and \citet{Gusakov:14}  provide enough detail to allow for a calculation of the Brunt-V\"{a}is\"al\"a frequency, $\mathcal{N}$, and thus g-modes.  Regarding the second point, in YW17 we solved the Newtonian equations of stellar structure in order to be consistent with our Newtonian treatment of the stellar oscillations and tidal driving (a relativistic treatment of the tides would significantly complicate the analysis and would not lead to substantially different results). However, $\approx 2 M_\odot$ Newtonian models do not reach high enough core densities to yield hyperons in GM1'B ($\rho \gtrsim 7 \times 10^{14}$ g/cm$^3$). We therefore construct GR background models which are more compact than Newtonian models and contain hyperons if $M\gtrsim 1.4 M_\odot$. For simplicity, however, we still solve for the g-modes and tidal driving using the Newtonian equations of stellar oscillation.   As we describe in Section \ref{sec:eigenmodes}, we carry out a partial check of the robustness of this hybrid approach by calculating some of the g-modes (but not tidal driving) using the GR equations of stellar oscillation.

We consider superfluid NS models with three different masses: $1.4 M_\odot$, $1.5 M_\odot$, and $1.6 M_\odot$.  The $1.4 M_\odot$ NS does not contain hyperons because its central density is too low whereas the $1.5M_\odot$ and $1.6 M_\odot$ hyperon star (HS) models both contain $\Lambda$ hyperons in the inner core. The $1.6 M_\odot$ HS mass is chosen such that the density of the inner core is high enough to contain $\Lambda$ hyperons but just slightly too low to  contain other hyperon species. In particular, a more massive NS in GM1'B would contain $\Xi^-$ and $\Xi^0$ hyperons which, while potentially interesting for tidal physics, would considerably complicate the analysis. Compared to the $1.6 M_\odot$ HS model, the $1.5 M_\odot$ HS model has a smaller mass fraction of hyperons in the inner core.  By comparing the results for the three models, we study how the presence and abundance of $\Lambda$ hyperons modify the g-mode oscillation spectrum and the dynamical tide GW phase shifts. 

\begin{table}
\begin{center}
\caption{\label{tab:bgConfig}Parameters of the background NS and HS models.}
\hspace*{-0.5cm}
\setlength{\tabcolsep}{4.5pt}
\begin{tabular}{cccccc}
\hline
$M$ [$M_\odot$] & $R$ [km]& $\rho_0$ [g$\ $ cm$^{-3}$] & $R_\Lambda$ [km]&$R_\mu$ [km] & $R_{\rm cc}$ [km] \\
\hline
1.4 & 13.7 & $6.5\times10^{14}$ & -- & 11.3 & 12.2 \\
1.5 & 13.6 & $7.1\times10^{14}$ & 3.3 & 11.5 & 12.3 \\
1.6 & 13.5 & $8.1\times10^{14}$ & 5.3  &11.6 & 12.4 \\
\hline
\end{tabular}
\end{center}
\end{table}

\begin{figure}
\includegraphics[angle=0,scale=.45]{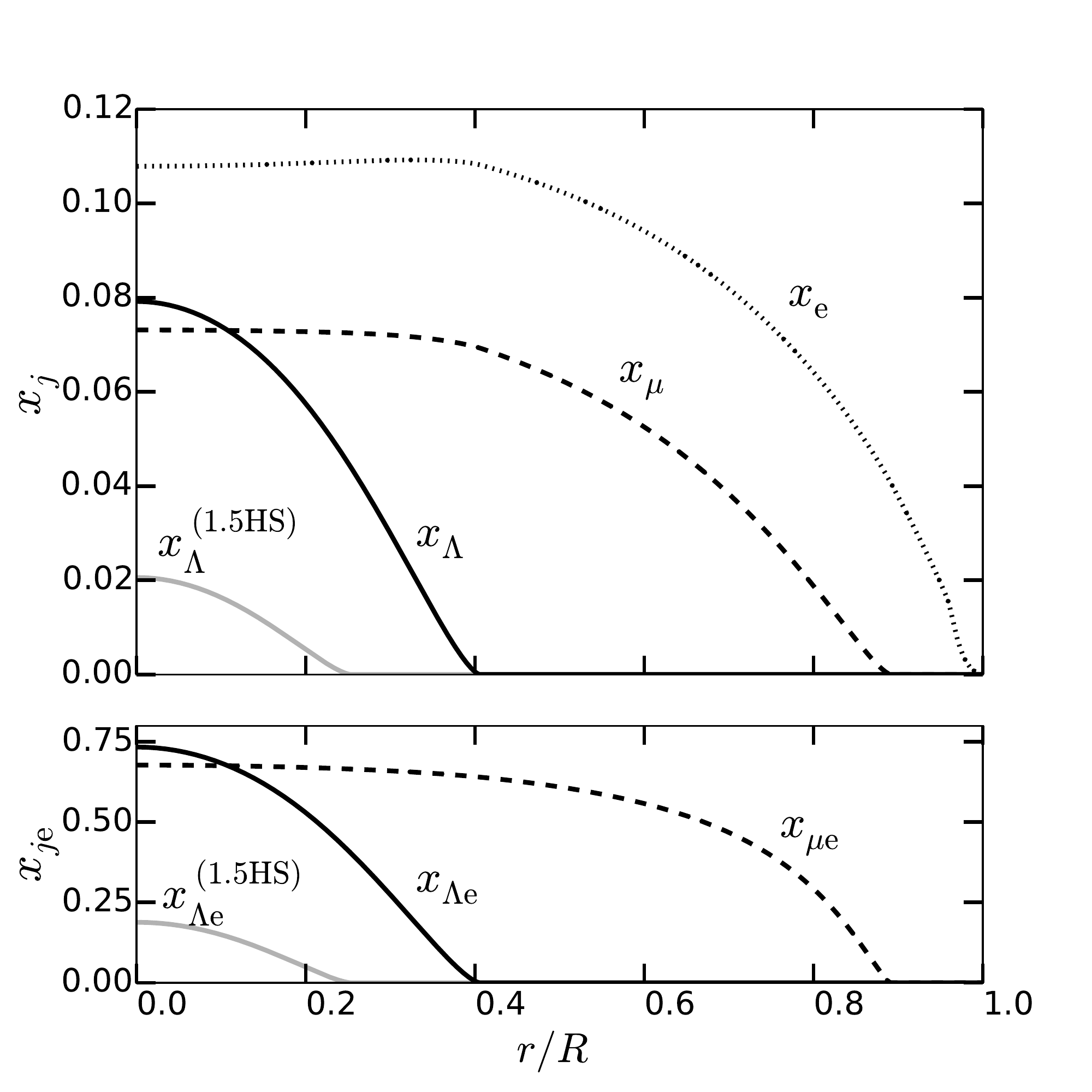}
\caption{Upper panel: number fraction  of electrons $x_{\rm e}$ (dotted line), muons $x_{\mu}$ (dashed line), and $\Lambda$ hyperons $x_\Lambda$ (solid black line) as a function of fractional radius $r/R$ for the $1.6 M_\odot$ HS model. For comparison, we also show $x_\Lambda$ for the $1.5 M_\odot$ HS model (solid grey line). Bottom panel: number of muons per electron $x_{\mu \rm e}$ (dashed line) and $\Lambda$ hyperons per electron $x_{\Lambda \rm e}$ (solid black line) for the $1.6 M_\odot$ HS model, and  $x_{\Lambda \rm e}$ for the $1.5 M_\odot$ HS model  (solid grey line).}
\label{fig:compGrad}
\end{figure}

\begin{figure*}
\includegraphics[angle=0,scale=.42]{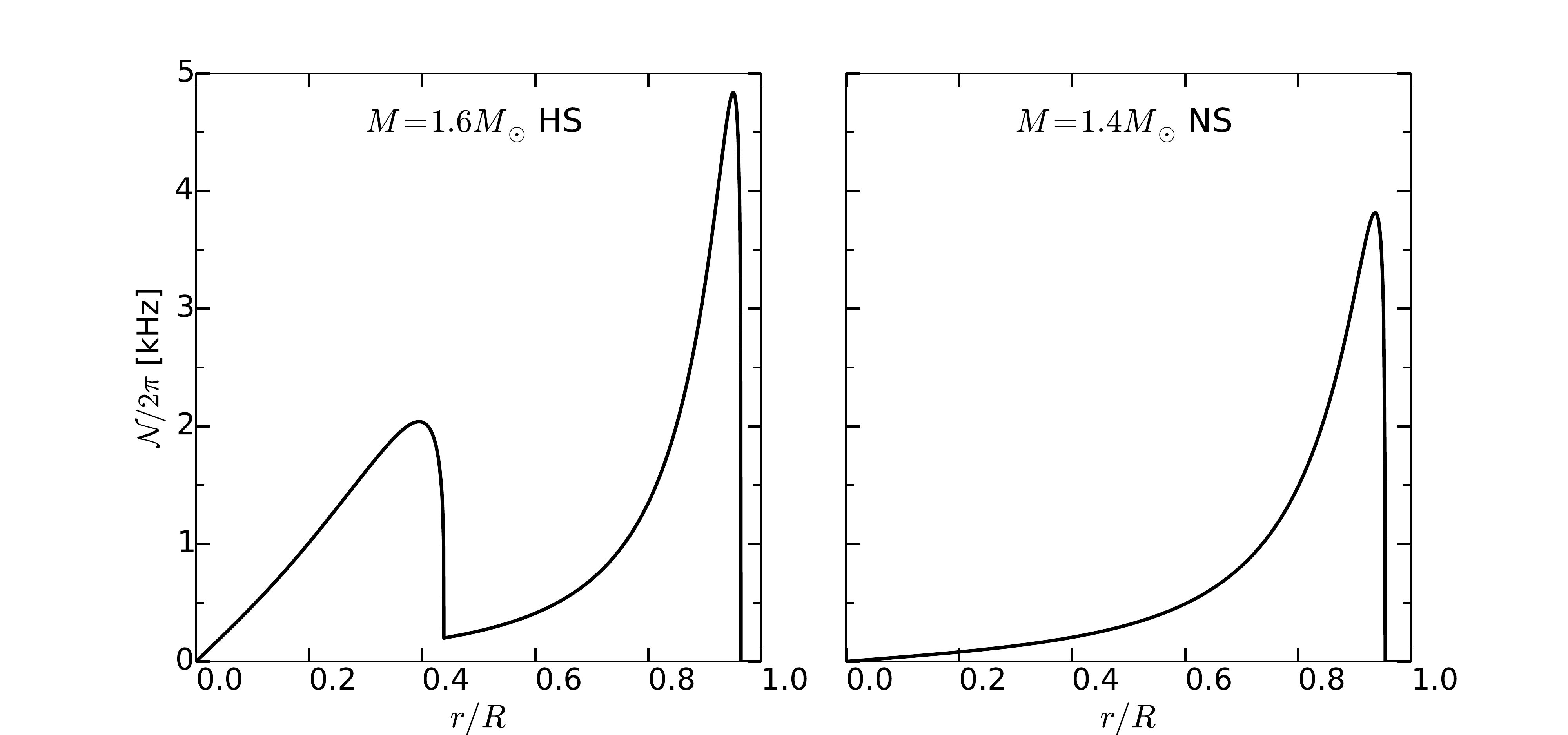}
\caption{Brunt-V\"{a}is\"al\"a frequency $\mathcal{N}/2\pi$ for the $1.6 M_\odot$ HS (left panel) and the $1.4 M_\odot$ NS (right panel).  Both models have a muonic contribution that peaks in the outer core. The HS has an additional contribution due to hyperons that peaks in the inner core.  Note that we evaluate the frequencies in the Newtonian limit (eq. \ref{eq:brunt}).}
\label{fig:BV}
\end{figure*}

The conditions for chemical equilibrium due to weak interactions are \citep{Dommes:16},
\begin{align}
&\mu_{\rm n}=\mu_{\rm p} + \mu_{\rm e}, \label{eq:chemEq1}\\
&\mu_{\rm n}=\mu_\Lambda, \label{eq:chemEq2}\\
&\mu_{\rm e} = \mu_\mu, \label{eq:chemEq3}
\end{align}
where $\mu_j$ is the chemical potential of particle species $j$,  with $j$ being either a neutron (n), proton (p), $\Lambda$ hyperon ($\Lambda$), electron (e), or muon ($\mu$). These conditions, along with the equations of stellar structure, determine the composition profile for each star.  Because of the presence of hyperons, the equation of state has four degrees of freedom instead of the three in YW17. In particular, we parametrize the thermodynamic relations in terms of the pressure $P$, the neutron chemical potential $\mu_{\rm n}$, the muon-to-electron ratio $x_{\mu \rm e}=x_\mu/x_{\rm e}$,  and the $\Lambda$-to-electron ratio $x_{\Lambda \rm e}=x_\Lambda / x_{\rm e}$, where $x_j=n_j/n_{\rm b}$ is the number fraction of species $j$ per baryon and $n_j$ is its number density. 

In Table \ref{tab:bgConfig}, we give the values of the following parameters for our three models: total mass $M$, radius $R$, central density $\rho_0$, radius below which $\Lambda$ hyperons are present $R_\Lambda$, radius below which muons are present $R_\mu$, and radius of the core-crust interface  $R_\textrm{cc}$ (defined as where the baryon density $n_{\rm b}=0.08 \textrm{ fm}^{-3}$).  In Fig. \ref{fig:compGrad}, we show the composition profile for the $1.6 M_\odot$ HS model and for comparison, $x_{\Lambda}(r)$ and $x_{\Lambda \rm e}(r)$ for the $1.5 M_\odot$ HS model.

Based on the discussion in YW17 (see their Section 2.1 and Appendix A),  in the Newtonian limit the Brunt-V\"{a}is\"al\"a (i.e., buoyancy) frequency is given by
\begin{align}
\label{eq:brunt}
\mathcal{N}^2=&-\frac{1-\epsilon_\textrm{n}}{1-x_\textrm{n}-\epsilon_\textrm{n}}\frac{g}{\rho}
\sum_{j=\mu, \Lambda}\left(\frac{\partial \rho}{\partial x_{j\rm e}}\right)\frac{\diff x_{j\rm e}}{\diff r},
\end{align}
where the partial derivative is evaluated by holding the three other thermodynamic parameters fixed ($P$, $\mu_{\rm n}$, and $x_{i\rm e}$ for $i\neq j$), $g(r)$ is the gravitational acceleration at radius $r$, and $\epsilon_{\rm n}$ is the superfluid entrainment function (YW17, see also \citealt{Prix:02}).  Numerical values for entrainment are provided in \cite{Gusakov:14} but using a different parameterization.

In the core, we evaluate the buoyancy  using equation (\ref{eq:brunt}).  In the crust, we follow YW17 and assume for simplicity that the crust is neutrally buoyant ($\mathcal{N}=0$); this does not significantly affect the core g-modes of interest here since the crust contains only a small fraction of the mass. In  Fig. \ref{fig:BV} we show $\mathcal{N}$ for the $1.4 M_\odot$ NS model (right panel) and the $1.6 M_\odot$ HS model (left panel). Whereas the $1.4 M_\odot$  model contains only a single $\mathcal{N}$ peak (due to the muon gradient $\diff x_{\mu \rm e}/\diff r$), the $1.6 M_\odot$ HS model  contains two $\mathcal{N}$ peaks (one due to the muon gradient $\diff x_{\mu \rm e}/\diff r$ and one at higher densities due to the $\Lambda$ hyperon gradient $\diff x_{\Lambda \rm e}/\diff r$).  As we show in the next section, this additional peak leads to a new type of g-mode, i.e., hyperonic g-modes. 

We assume that $x_{\mu \rm e}$ and $x_{\Lambda \rm e}$ do not vary during oscillations of the normal fluid; i.e., the composition  is ``frozen" and thus the perturbed fluid element is out of chemical equilibrium.  \citet{Reisenegger:92} consider the timescale for the  proton fraction $x_{\rm p}$ in a normal fluid NS to relax towards chemical equilibrium due to Urca processes ($x_p$ is the source of buoyancy in a normal fluid NS).  They show that for even a moderately warm NS, the relaxation timescale is much longer than the oscillation period of low-order g-modes and therefore, to a very good approximation, $x_{\rm p}$ is frozen within the fluid element.  Similarly, to check whether $x_{\Lambda \rm e}$ is frozen, we must consider the direct Urca process $\Lambda \to {\rm p} + {\rm L} +\bar{\nu}_{\rm L}$, where the lepton in the reaction $\textrm{L} = \rm{e}$ or $\mu$.  In Appendix \ref{appendix:urca} we show that the corresponding relaxation timescale is much longer than the oscillation period of low order hyperonic g-modes and therefore the assumption of frozen composition should also hold for these modes.

\section{EIGENMODES OF A SUPERFLUID HYPERON STAR}\label{sec:eigenmodes}
The oscillation equations of a superfluid HS are similar to those of a superfluid NS and can be written as (\citealt{Prix:02}, YW17)
\begin{align}
&\vect{\nabla}\cdot (\rho_\textrm{c} \vect{\xi}_\textrm{c}) + \delta \rho_\textrm{c}=0, \label{eq:mass1}\\
&\vect{\nabla}\cdot(\rho_\textrm{n} \vect{\xi}_\textrm{n}) + \delta \rho_\textrm{n}=0, \label{eq:mass2}\\
&\sigma^2\left[\vect{\xi}_\textrm{c} - \epsilon_\textrm{c}(\vect{\xi}_\textrm{c}-\vect{\xi}_\textrm{n}) \right] = \vect{\nabla}\left(\delta \tilde{\mu}_\textrm{c} + \delta \Phi\right), \label{eq:mom1}\\
&\sigma^2\left[\vect{\xi}_\textrm{n} + \epsilon_\textrm{n}(\vect{\xi}_\textrm{c}-\vect{\xi}_\textrm{n}) \right] = \vect{\nabla}\left(\delta \tilde{\mu}_\textrm{n} + \delta \Phi\right), \label{eq:mom2}\\
&\nabla^2\delta \Phi=4\pi G (\delta \rho_\textrm{c} + \delta \rho_\textrm{n}), \label{eq:poisson}
\end{align} 
where we assume that the perturbed quantities have a time dependence $e^{\rm{i}\sigma t}$, $\delta \mathcal{Q}(\vect{x})$ denotes the Eulerian perturbation of a quantity $\mathcal{Q}$ at position $\vect{x}$,  subscript `n' denotes the neutron superfluid flow, and subscript `c' denotes the normal fluid flow (consisting of the charged particles and the $\Lambda$ hyperons; we continue to use a subscript c in order to match the notation used in \citet{Prix:02} and YW17).  The other quantities are the mass densities $\rho_\textrm{c(n)}$ (the total density $\rho=\rho_{\rm c}+\rho_{\rm n}$), the perturbed specific chemical potentials $\delta \tilde{\mu}_\textrm{c(n)}$, the Lagrangian displacements $\vect{\xi}_\textrm{c(n)}$, the perturbed gravitational potential $\delta \Phi$, and the entrainment function $\epsilon_{\rm c}=\epsilon_{\rm n}\rho_{\rm n}/\rho_{\rm c}$  (see YW17 for further details).

Although the oscillation equations take a simple form when written in terms of $\vect{\xi}_\textrm{c}$ and $\vect{\xi}_\textrm{n}$, it is more convenient to express the tidal excitation of modes in terms of the mass-averaged flow $\vect{\xi}_+$ and the difference flow $\vect{\xi}_-$, where
\begin{align}
&\vect{\xi}_+=\frac{1}{\rho}\left(\rho_\textrm{c}\vect{\xi}_\textrm{c} + \rho_\textrm{n}\vect{\xi}_\textrm{n}\right),  \label{eq:xi_plus}\\
&\vect{\xi}_-=\left(1-\epsilon_\textrm{n}-\epsilon_\textrm{c}\right)\left(\vect{\xi}_\textrm{c}-\vect{\xi}_\textrm{n}\right).
\label{eq:xi_minus}
\end{align}
When solving the oscillation equations, we choose ($\vect{\xi}_+$, $\vect{\xi}_{\rm n}$, $\delta P$, $\delta \tilde{\mu}_\textrm{n}$, $\delta \Phi$) to be our  independent variables. We can then use equations (\ref{eq:xi_plus}) and (\ref{eq:xi_minus}) to calculate $\vect{\xi}_{\rm c}$ and $\vect{\xi}_{-}$. Since the Lagrangian perturbations to $x_{\mu \rm e}$ and $x_{\Lambda \rm e}$ vanish for a frozen composition, we can use the chain rule to express the density perturbation as

\begin{align}
\delta \rho_\textrm{c} &=  \left(\frac{\partial \rho_\textrm{c}}{\partial P}\right)\delta P + \left(\frac{\partial \rho_\textrm{c}}{\partial \tilde{\mu}_\textrm{n}}\right)\delta \tilde{\mu}_\textrm{n}%\nonumber \\
%&\quad - \left(\frac{\partial \rho_\textrm{c}}{\partial x_{\mu \rm e}}\right) \vect{\nabla}x_{\mu \rm e}\cdot\vect{\xi}_\textrm{c}^r - \left(\frac{\partial \rho_\textrm{c}}{\partial x_{\Lambda \rm e}}\right)\vect{\nabla} x_{\Lambda \rm e}\cdot \vect{\xi}_\textrm{c}^r,&
- \sum_{j=\mu,\Lambda} \left(\frac{\partial \rho_\textrm{c}}{\partial x_{j\rm e}}\right) 
%\vect{\nabla}x_{\mu j}\cdot\vect{\xi}_\textrm{c}^r,
\frac{\diff x_{j\rm e}}{\diff r}\xi^r_{\rm c},
\end{align}
and similarly for $\delta \rho_\textrm{n}$, where $\xi^r_{\rm c}(r)$ denotes the radial dependence of the radial component of $\vect{\xi}_{\rm c}$ (the angular dependence is given by the spherical harmonic function $Y_{l m}(\theta,\phi)$ of degree $l$ and order $m$). As we show in YW17 (see also \citealt{Lindblom:94, Andersson:04}), the operator $\mathcal{L}$ corresponding to equations (\ref{eq:mass1})-(\ref{eq:poisson}) is Hermitian with respect to the inner product
\begin{equation}
\left\langle 
\begin{bmatrix}
 \vect{\xi}_{+} \\ \vect{\xi}_{-}
\end{bmatrix},
\begin{bmatrix}
\vect{\xi}_{+}' \\
\vect{\xi}_{-}' 
\end{bmatrix}\right\rangle
=\int \diff^3 x 
\begin{bmatrix}
 \vect{\xi}_{+}^\ast & \vect{\xi}_{-}^\ast
\end{bmatrix}
\begin{bmatrix}
\rho & 0 \\
0 & \tilde{\rho}
\end{bmatrix}
\begin{bmatrix}
\vect{\xi}_{+}' \\
\vect{\xi}_{-}' 
\end{bmatrix},
\label{eq:innerProduct}
\end{equation}
where 
\begin{equation}
\tilde{\rho}=\frac{\rho_\textrm{c}\rho_\textrm{n}}{(1-\epsilon_\textrm{n} - \epsilon_\textrm{c})\rho}.
\end{equation}
The set of eigenmodes thus forms an orthonormal base. We use the same boundary conditions as YW17 and  normalize the modes such that 
\begin{equation}
\sigma_a^2\left\langle 
\begin{bmatrix}
 \vect{\xi}_{a+} \\ \vect{\xi}_{a-}
\end{bmatrix},
\begin{bmatrix}
\vect{\xi}_{b+} \\
\vect{\xi}_{b-} 
\end{bmatrix}\right\rangle
= E_0\delta_{ab},
\label{eq:SForthonormal}
\end{equation}
where $\sigma_a$ is the eigenvalue of mode $a$ and $E_0=GM^2/R$.

\begin{figure}
\hspace{-0.2cm}
\includegraphics[angle=0,scale=.6]{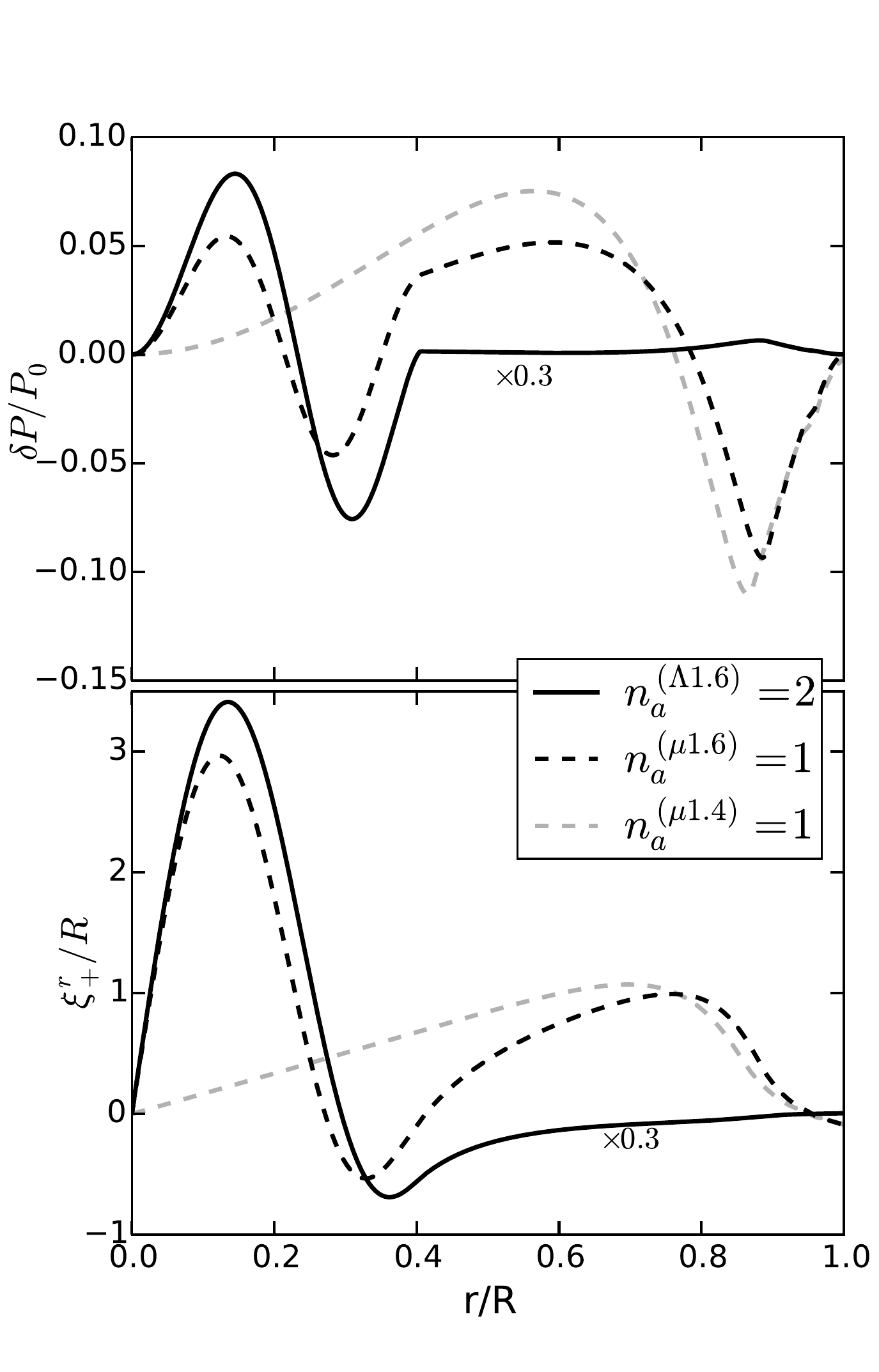}
\caption{Structure of the second hyperonic g-mode $n_a^{(\Lambda1.6)}=2$ (solid black lines) and the first muonic g-mode $n_a^{(\mu1.6)}=1$  (dashed black lines) of the $1.6 M_\odot$ HS and the first g-mode $n_a^{(\mu1.4)}=1$ of the 1.4 $M_\odot$ NS (dashed grey  lines). All three modes have spherical degree $l_a=2$. The upper panel shows the Eulerian perturbation of the pressure $\delta P$ in units of the central pressure $P_0$.  The lower panel shows the radial component of the mass-averaged Lagrangian displacement $\xi^r_+$ in units of $R$. All quantities are normalized according to Equation (\ref{eq:SForthonormal}); for display purposes, those corresponding to the $n_a^{(\Lambda1.6)}=2$ mode are multiplied by an additional factor of 0.3.}
\label{fig:modeStruct}
\end{figure}

As  \cite{Dommes:16} first showed, the core g-modes of an HS can be classified into two types: `hyperonic' g-modes and `muonic' g-modes.  The hyperonic g-modes are primarily supported by the $\Lambda$  hyperon gradient and are concentrated in the inner core, while the muonic g-modes  are supported by a combination of the $\Lambda$ hyperon and muon gradients and span both the inner and outer core. In Fig. \ref{fig:modeStruct} we show the structure of the second hyperonic g-mode $n_a^{(\Lambda1.6)}=2$  and the first muonic g-mode $n_a^{(\mu1.6)}=1$  of our $1.6 M_\odot$ HS model.\footnote{We use $n_a^{(\Lambda1.6)}$ and $n_a^{(\mu1.6)}$ to label the sequential order of each type of g-mode, with $n_a^{(\Lambda1.6)}=1$ and $n_a^{(\mu1.6)}=1$ corresponding to the highest frequency hyperonic g-mode and muonic g-mode, respectively. They do not necessarily correspond to the mode's radial order, i.e., the number of radial nodes.}  In the superscript, $\Lambda1.6$ ($\mu1.6$) stands for the hyperonic (muonic) modes of the 1.6 $M_\odot$ HS model. We will use this labeling convention throughout the paper. For comparison, we also show the first g-mode of our $1.4 M_\odot$ NS model. All three g-modes have angular degree $l_a=2$ and thus  couple to the $l=2$ harmonic of the tide.

\begin{figure}
\includegraphics[angle=0,scale=.4]{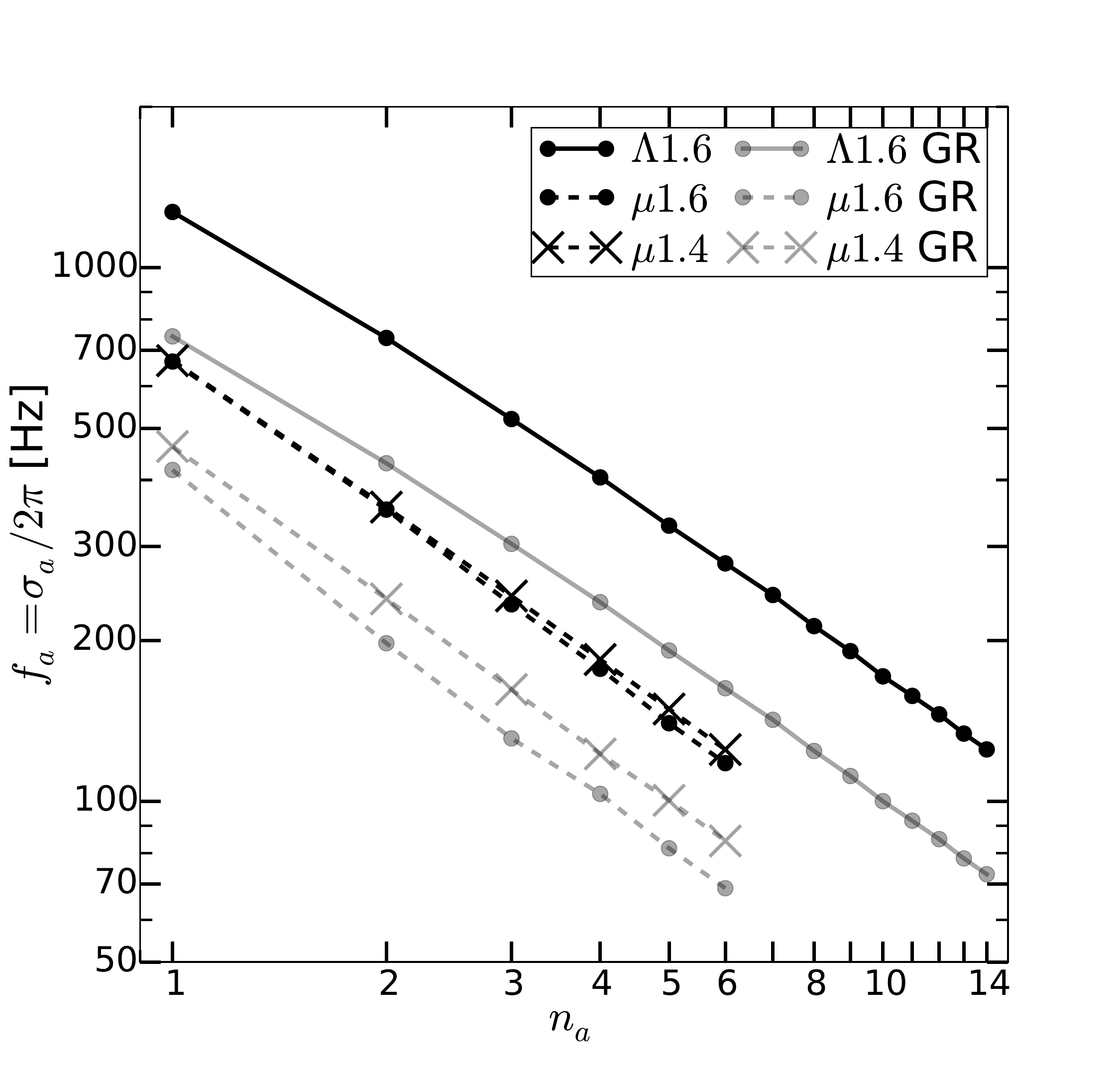}
\caption{Eigenfrequencies $f_a=\sigma_a/2\pi$ of $l_a=2$  g-modes calculated with the Newtonian oscillation equations (black lines) and the GR oscillation equations (grey lines).  For the $1.6 M_\odot$ HS model, we show the hyperonic modes with filled circles connected by solid lines and the muonic modes with filled circles connected by dashed lines.  For the $1.4M_\odot$ NS model, which only have muonic g-modes, we use crosses connected by dashed lines.}
\label{fig:modeFreq}
\end{figure}

In Fig. \ref{fig:modeFreq} we show the eigenfrequencies $f_a=\sigma_a/2\pi$ of the first several $l_a=2$ g-modes of our $1.6 M_\odot$ HS and $1.4 M_\odot$ NS models. To guide the eye, we use straight lines to connect each of the hyperonic modes (solid) and, separately, each of the muonic modes (dashed).  As we describe in Section \ref{sec:bgModel}, in order to build  HSs without complicating the dynamical tide calculation, we solve the TOV equations to construct the background models but the Newtonian equations to calculate the g-modes and tidal driving. We partially assess the impact of this hybrid approach by redoing the calculation of the g-modes using the GR oscillation equations.  For this calculation, we ignore the gravitational perturbations (i.e., we adopt the Cowling approximation) and solve the superfluid GR oscillation equations (see, e.g.,  \citealt{Dommes:16, Passamonti:16}).  We find that our hybrid approach overestimates the g-mode eigenfrequencies by $\simeq 70 \%$ (see grey lines in Fig. \ref{fig:modeFreq}).  For example, the highest frequency $l_a=2$ hyperonic g-mode in the $1.6 M_\odot$ HS has a frequency of $f_a^{\rm(GR)}=743\textrm{ Hz}$ in the fully relativistic calculation (as seen by an observer at infinity) compared to $f_a=1271\textrm{ Hz}$ in the hybrid calculation. Nevertheless, as we show in Section \ref{sec:tides}, the dynamical tide phase shift is independent of the eigenfrequency $f_a$ (in our normalization) and is therefore unaffected by the overestimate of $f_a$. There is still an error due to our hybrid calculation of the tidal coupling strength, but we argue in Section \ref{sec:tides} that this should not be too significant. 

Consistent with the asymptotic properties of high-order g-modes \citep{Aerts:10}, we find that for modes with $n_a^{(j)} \gtrsim \textrm{ few}$, the GR oscillation equations yield $l_a=2$ eigenfrequencies that are well approximated by $f_a^{(j, \textrm{GR})}={f_0^{(j)}}/{n_a^{(j)}}$,
where $f_0^{(j)}=\{1020, 410, 460, 500, 510\}\textrm{ Hz}$ for $j=\{\Lambda1.6, \mu1.6, \Lambda1.5, \mu1.5, \mu1.4\}$, respectively.
We find that the characteristic frequency $f_0^{(\Lambda)}$ of the hyperonic modes scales almost linearly with the size of the hyperonic core $R_\Lambda$.   This is because $f_0^{(\Lambda)} \propto \int_0^{R_\Lambda} \mathcal{N} d\ln r$ and for $r\lesssim R_\Lambda$,  the density $\rho\simeq \textrm{ constant}$ and $\mathcal{N} \propto r$.

\section{TIDAL DRIVING AND PHASE SHIFT of the gravitational waveform}\label{sec:tides}

Following YW17 (see also \citealt{Lai:94, Weinberg:12}), we solve for the resonant tidal excitation of g-modes by expanding the tidal displacement field as
\begin{equation} 
\begin{bmatrix}
\vect{\xi}_+ (\vect{x},t)\\
\vect{\xi}_-(\vect{x},t)
\end{bmatrix}
=\sum_a b_a(t)
\begin{bmatrix}
\vect{\xi}_{a+} (\vect{x})\\
\vect{\xi}_{a-}(\vect{x})
\end{bmatrix},
\label{eq:mode_decomp}
\end{equation}
where the subscript $a=\{n_a,\ l_a,\ m_a\}$ denotes a specific eigenmode of the NS and $b_a(t)$ is the time-dependent, dimensionless amplitude of mode $a$ (a mode with amplitude $|b_a|=1$ has energy $E_0=GM^2/R$). Inserting this expansion into the linear superfluid oscillation equations (\ref{eq:mass1})-(\ref{eq:poisson}) and including the time-dependent  tidal potential ($\delta \Phi \rightarrow \delta \Phi +U$) yields the mode amplitude equation
\begin{equation}
\ddot{b}_a + \sigma_a^2 b_a =\sigma_a^2 U_a(t),
\label{eq:amp_eqn}
\end{equation}
where the tidal driving coefficient 
\begin{equation}
U_a(t) = \frac{M'}{M}\sum_{l\ge2,m} W_{lm} Q_{alm} \left(\frac{R}{D(t)}\right)^{l+1} e^{-im\psi(t)}.
\label{eq:tide_coefficient}
\end{equation}
Here $M'$ is the mass of the companion,  $lm$ are harmonics of the tidal potential, the coefficients $W_{lm}=4\pi(2l+1)^{-1}Y_{lm}(\pi/2,0)$ are of order unity for the dominant harmonics \citep{Press:77}, $D(t)$ is the orbital separation, and $\psi(t)$ is the orbital phase.  

The time-independent, dimensionless tidal coupling coefficient (sometimes referred to as the tidal overlap integral)
\begin{equation}
Q_{alm} =\frac{1}{MR^l}\int \diff^3 x \rho \, \vect{\xi}_{a+}^\ast \cdot\vect{\nabla}\left(r^l Y_{lm}\right).
\label{eq:Qform_orig}
\end{equation}
By angular momentum conservation, $Q_{alm}$ is non-zero only if $l_a=l$ and $m_a=m$. Based on our numerical calculations, we find that the $Q_{alm}$ of our different models and mode types are approximately given by 

\begin{equation}
|Q_{alm}^{(j)}| = Q_0 n_a^{-\alpha},
\label{eq:Qvsn}
\end{equation}
where
\begin{align}
&\alpha\approx 2.5 \hspace{0.25cm} \textrm{and} \nonumber \\
&Q^{(j)}_0=\{1.7,\,3.6,\,1.6,\,3.5,\,4.0\}\times 10^{-3} ,
\label{eq:Qcoefs}
\end{align}
for $j=\{\Lambda1.6, \mu1.6, \Lambda1.5, \mu1.5, \mu1.4\}$, respectively.

Although these values are based on the hybrid calculation (GR background and Newtonian oscillations), we expect the  error due to this inconsistency to be relatively small. As an approximate measure of the correction,  we computed $|Q_{alm}|$ for the $1.4 M_\odot$ NS model but with the background constructed using the Newtonian structure equations rather than the TOV equations (we could not perform such a test for the HS model because a $2.0 M_\odot$ Newtonian model does not contain hyperons).  We find that $Q_{alm}$ changes by at most $5\%$, which suggests that our hybrid model gives a reasonably accurate estimate of the tidal coupling strength.  Indeed, given that  $GM/Rc^2\approx20\%$, we do not expect relativistic corrections to be much bigger than a few tens of percent.

We numerically solve the amplitude  equation (\ref{eq:amp_eqn}) of the resonantly driven g-modes and find that the result is in good agreement with the analytic solution given by the stationary-phase approximation \citep{Lai:94, Reisenegger:94}.  The GW phase shift induced by the resonant excitation of each mode is therefore approximately given by (YW17)
\begin{equation}
\delta \phi_a=-\frac{5\pi^2}{2048}k'\left(\frac{GM}{Rc^2}\right)^{-5}\sum_{m=\pm2}|Q_{a2m}|^2,
\label{eq:dphia_genl}
\end{equation}
where $k'=2/[q(1+q)]$, $q=M'/M$, and here the subscript $a$ already accounts for the contributions from both the $m_a = 2$ and $m_a = -2$ modes. Since $\delta \phi_a$ depends on $f_a$ only through $|Q_{alm}|$, and we expect that our hybrid approach accurately estimates $|Q_{alm}|$ to within a few tens of percent, our estimate of $\delta \phi_a \propto |Q_{alm}|^2$ should  be accurate to within a factor of order unity.

In Table \ref{tab:delPhi} we list $f^{\rm(GR)}_a$ (see grey lines in Figure \ref{fig:modeFreq}) and $\delta \phi_a$ for the three-lowest order $l_a=2$ g-modes of each type, for each stellar model.  Numerically, we find that summing over all of the resonantly excited modes yields a total cumulative phase shift 
\begin{align}
\label{eq:dphi_HS}
&\delta \phi^{\textrm{(1.6HS)}}
%=\Delta \phi^{(\Lambda)}+\Delta \phi^{(\mu)} 
=-4.3 \times 10^{-3}k', \\
\label{eq:dphi_HS15}
&\delta \phi^{\textrm{(1.5HS)}}
%=\Delta \phi^{(\Lambda)}+\Delta \phi^{(\mu)} 
=-6.7 \times 10^{-3}k', \\
\label{eq:dphi_NS}
&\delta \phi^{\textrm{(1.4NS)}}
%=\Delta \phi^{(1.4)}
=-9.9 \times 10^{-3}k',
\end{align}
where $\delta \phi^{\textrm{(HS)}}$ accounts for both the  hyperonic modes 
and muonic modes. 

As can be gleaned from Table \ref{tab:delPhi} [also eqs. (\ref{eq:Qvsn}) - (\ref{eq:dphia_genl})], the total phase shift is dominated by the contributions of the $n_a=1$ g-modes (the modes with the highest frequency).  The phase shifts due to muonic modes are insensitive to whether hyperons are present in the core and decrease with increasing HS/NS mass. In the HS models, the cumulative phase shift due to the muonic g-modes is about five times larger than that due to the hyperonic g-modes.  
Moreover, comparing the $1.6 M_\odot$ HS model to the $1.5 M_\odot$ HS model, we find that the phase shift due to the hyperonic modes is relatively insensitive to the size of the hyperon core $R_\Lambda$, in contrast to the eigenfrequency spectrum which is shifted to lower frequencies as the mass of the hyperon core decreases (see discussion at end of Section \ref{sec:eigenmodes}).

\begin{table}
\begin{center}
\caption{\label{tab:delPhi} Eigenfrequencies $f_a^{\rm(GR)}$ and gravitational waveform phase shift $\delta \phi_a$ for the three-lowest order $l_a=2$ g-modes for each stellar model. The format is $\left\{f_a/100\textrm{ Hz}, \,|\delta \phi_a|/\textrm{rad}\right\}$. }
\hspace*{-0.5cm}
\setlength{\tabcolsep}{0.18cm}
\begin{tabular}{cccc}
\hline
 $j$ &$n_a^{(j)}=1$& $n_a^{(j)} =2$ & $n_a^{(j)}=3$  \\
\hline
$\Lambda1.6$ & \{7.4, \, 5.5e-4\} &\{4.3, \, 3.4e-4\} & \{3.0, \, 3.6e-7\}\\
$\mu1.6$ & \{4.1, \, 3.4e-3\} & \{2.0, \, 1.6e-5\} & \{1.3, \, 1.2e-5\} \\
$\Lambda1.5$ & \{4.1, \, 1.1e-3\} & \{2.5, \, 1.6e-7\} & \{1.4, \, 6.6e-7\} \\
$\mu1.5$ & \{4.5, \, 5.4e-3\} & \{2.4, \, 6.6e-6\} & \{1.6, \, 2.1e-5\}\\
$\mu1.4$ & \{4.6, \, 9.8e-3\} & \{2.4, \, 3.6e-7\} & \{1.6, \, 3.0e-5\}\\
\hline
\end{tabular}
\end{center}
\end{table}

\section{DETECTABILITY OF THE MODE RESONANCES}
\label{sec:detectability}
In this section we estimate the detectability of the dynamical tide phase shift using the Fisher matrix formalism \citep{Cutler:94}.  In Section \ref{sec:detectability_model}, we describe how we model the GW signal and the detectability of the resonances based on single events and multiple stacked events.    In Section \ref{sec:detectability_results}, we present our detectability estimates for Advanced LIGO and proposed third generation GW detectors. 

\subsection{Modeling the detectability}
\label{sec:detectability_model}
Following \citet{Cutler:94}, if we assume a strong GW strain signal $h(t)$ and Gaussian detector noise, then the signal parameters $\theta^i$ have a probability distribution $p(\theta^i)\propto \exp\left[-(1/2)\Gamma_{ij} \diff \theta^i\diff \theta^j\right]$, where $\diff \theta^i=\theta^i-\hat{\theta}^i$ is the difference between the parameters and their best fit values $\hat{\theta}^i$ and 
\begin{equation}
%\Gamma_{ij} = \left(\frac{\partial \tilde{h}}{\partial \theta^i}\Bigg{|}\frac{\partial \tilde{h}}{\partial \theta^j}\right)
\Gamma_{ij} = \left(\frac{\partial h}{\partial \theta^i}\Bigg{|}\frac{\partial h}{\partial \theta^j}\right)
\end{equation}
is the Fisher information matrix.
 The parentheses denote the inner product 
\begin{equation}
%\left(h_1| h_2\right)=2\int_0^{2f_{\rm ISCO}}\frac{\tilde{h}_1(f)^\ast \tilde{h}_2(f) + \tilde{h}_1(f) \tilde{h}_2^\ast(f)}{S_n(f)}\diff f, 
\left(h_1| h_2\right)=2\int_0^{\infty}\frac{\tilde{h}^\ast_1(f) \tilde{h}_2(f) + \tilde{h}_1(f) \tilde{h}_2^\ast(f)}{S_n(f)}\diff f, 
\end{equation}
where $\tilde{h}_1$ and $\tilde{h}_2$ are the Fourier transforms of $h_1$ and $h_2$ and $S_n(f)$ is the strain noise power spectral density. The signal-to-noise ratio (SNR) of a signal $h$ is given by $\textrm{SNR} = (h|h) ^{1/2}$ and the root-mean-square (rms) measurement error  in $\theta^i$ is given by a diagonal element of the inverse Fisher matrix 
\begin{equation}
%\Delta\left(\theta^i\right) = \sqrt{<(\diff \theta^i)^2>} = \sqrt{\Sigma^{ii}}, \hspace{0.3cm} \text{with } \vect{\Sigma}=\vect{\Gamma}^{-1},
\Delta \theta^i  = \sqrt{\left(\vect{\Gamma}^{-1}\right)^{ii}}.
\label{eq:rms_error}
\end{equation}
If the rms error in, say, the phase shift due to a particular resonance $\Delta(\delta \phi_a)$ is smaller than $\delta \phi_a$, then that phase shift is detectable.

The above analysis is based on the detection of a single inspiral event. We can roughly estimate how stacking multiple events would affect the measurement accuracy by using the  approach given in \citet{Markakis:10}; a more precise estimate would require a fully Bayesian investigation and is beyond the scope of this paper (see, e.g., the recent studies of the detectability of the equilibrium tide phase shift by \citealt{DelPozzo:13, Lackey:15, Agathos:15}).  The estimate assumes that the events are homogeneously distributed in a sphere of effective radius   $d^{\rm eff}_{\rm max}$ and that the error in each parameter $\Delta \theta$ scales linearly with effective distance $d^{\rm eff}< d^{\rm eff}_{\rm max}$.\footnote{The effective distance $d^{\rm eff}$ is defined as the distance obtained by averaging over a uniform source orientation for an event with a given SNR. It is related to the horizon distance $d^{\rm hor}$, the distance assuming an optimal source orientation, as $d^{\rm eff} =d^{\rm hor}/2.3$ (see, e.g., Appendix D of \citealt{Allen:12}).} Then, \citet{Markakis:10} show that the rms error averaged over all of the events is 
\begin{equation}
\langle{\Delta \theta}\rangle \approx\frac{\Delta \theta_{\rm ref}}{d^{\rm eff}_{\rm ref}\sqrt{4\pi \mathcal{R} P_{\rm obs}d^{\rm eff}_{\rm max}}}, \label{eq:stack_events}
\end{equation}  
where $\Delta \theta_{\rm ref}$ is the rms error of parameter $\theta$ from a single reference event at effective distance $d_{\rm ref}^{\rm eff}$, $\mathcal{R}$ is the event rate per unit time per unit volume, and $P_{\rm obs}$ is the total observation period. In the above calculation we ignore the cosmological expansion and assume that the event rate is constant. 

Similar to \citet{Balachandran:07}, who also studied the detectability of mode resonances (see also \citealt{Flanagan:07}), we assume that the frequency-domain GW signal has the form
\begin{equation}
\tilde{h}(f)  = \mathcal{A} f^{-7/6} e^{i\Psi(f)},
\label{eq:h_of_f}
\end{equation}
where the amplitude \citep{Maggiore:10} 
\begin{equation}
\mathcal{A}=\left(\frac{5}{24\pi^{4/3}}\right)^{1/2}\left(\frac{G\mathcal{M}}{c^3}\right)^{-5/6}\frac{\mathcal{K}}{d}.
\end{equation}
Here $\mathcal{M}=[(M M')^{3}/(M+M')]^{1/5}$ is the chirp mass, $d$ is the distance to the source, and $\mathcal{K}$ is the antenna response (for an optimally oriented source $\mathcal{K}=1$). Our model of the phase evolution $\Psi(f)$ accounts for the zeroth order post-Newtonian point-particle contribution and for the resonant tidal excitation of individual g-modes. We ignore higher order post-Newtonian terms, spin, the equilibrium tide, and nonlinear tidal effects (such as those considered in \citealt{Essick:16}), and assume that the signal shuts off at the gravitational wave frequency corresponding to the inner-most stable circular orbit
%$f=(6^{3/2} \pi G(M+M')/c^3)^{-1}$
$f=2f_{\rm ISCO} \simeq 1.6\times10^3 \textrm{Hz } [2.8 M_\odot/(M+M')]$ \citep{Cutler:94,Poisson:95}.
In Appendix \ref{appendix:phase} we show that under these assumptions,
\begin{equation}
\Psi(f)
= \Psi_{\rm pp}(f) - \sum_a\left(1-\frac{f}{f_a}\right) \delta \phi_a \Theta\left(f-f_a\right),
\end{equation} 
where, by the stationary-phase approximation,  the point-particle phase is \citep{Cutler:94}
\begin{equation}
\Psi_{\rm pp}(f) = 2\pi f t_c -\phi_c-\frac{\pi}{4} +\frac{3}{4}\left(\frac{8\pi G\mathcal{M}f}{c^3}\right)^{-5/3},
\label{eq:phi_pp}
\end{equation}
$\delta \phi_a$ is the phase shift due to the tidal resonance with a mode $a$ with eigenfrequency $f_a$,  and $\Theta\left(f-f_a\right)$ is the Heaviside step function.
Here $t_c$ and $\phi_c$ are constants of integration that set a reference time and phase.  
The duration of each resonance is, in general, much shorter than the orbital decay timescale due to radiation reaction (their ratio is $\simeq 0.1\times[(\mathcal{M}/1.2M_\odot)(f/500 \text{ Hz})]^{5/6}$; see \citealt{Lai:94, Flanagan:07,Balachandran:07}).  We therefore model the resonance as an instantaneous process, which should be a good approximation since nearly all the g-modes we consider have $f_a^{\rm (GR)} < 500\textrm{ Hz}$ (the one exception is the $n_a=1$ hyperonic g-mode of the $1.6 M_\odot$ HS model which has $f_a^{\rm (GR)}=740\textrm{ Hz}$).

Since the phase shift is dominated by the resonant excitation of the lowest order mode, for simplicity we do not sum over the modes in our waveform model but instead just consider the phase shift due to a single mode.  Our model therefore depends on 6 parameters: $\mathcal{A}$, $\mathcal{M}$, $t_c$, $\phi_c$, $f_a$, and $\delta \phi_a$.  \cite{Flanagan:07} also consider the effect of phase shifts on the GW signal, although their model is written in terms of the phase of the time-domain waveform $\phi(t)$ (their equation 1.9) rather than the frequency-domain waveform $\Psi(f)$.  In Appendix \ref{appendix:phase} we show that the two treatments are consistent.  

Before examining the numerical results, note that 
\begin{align}
&\frac{\partial \tilde{h}}{\partial (\delta \phi_a)} = - \textrm{i} \left(1-\frac{f}{f_a}\right) \Theta\left(f-f_a\right) \tilde{h}, \label{eq:dh_dphi_a}\\
&\frac{\partial \tilde{h}}{\partial f_a}= -\textrm{i} \delta \phi_a \frac{f}{f_a^2}\Theta \left(f-f_a\right)\tilde{h} \label{eq:dh_df_a}.
\end{align}
Note that the $(1-f/f_a)$ factor eliminates the $\delta$-function at $f=f_a$ stemming  from the derivative of $\Theta$. By equation (\ref{eq:rms_error}), we see that $\Delta (\delta \phi_a)$ and $\Delta f_a$ both vary linearly with distance (and thus inversely with SNR) and that $\Delta (\delta \phi_a)$ is independent of  $\delta \phi_a$ whereas $\Delta f_a \propto (\delta \phi_a)^{-1}$. Conceptually, these last two properties reflect the fact that the measurability of phase shifts $\Delta (\delta \phi_a)$ is mostly determined by how much SNR accumulates before and after the resonance (which is independent of $\delta \phi_a$ itself), whereas the larger $\delta \phi_a$ is, the easier it is to localize the frequency of the resonance (and thus the smaller $\Delta f_a$ is). Lastly, because the dominating g-modes are excited at frequencies higher than the most sensitive band of ground-based GW detectors ($\approx 70 \textrm{ Hz}$), larger $f_a$ have larger $\Delta (\delta \phi_a)$ and $\Delta (f_a)$ and thus worse detectability. This is different from the results in \cite{Balachandran:07} who found that increasing $f_a$ would make the detection easier, as \cite{Balachandran:07} considered modes that were excited at frequencies lower than the most sensitive band.

\subsection{Detectability with second and third generation detectors}
\label{sec:detectability_results}

\begin{table}
\begin{center}
\caption{\label{tab:detectability}Threshold distance $d_{\rm th}^{\rm hor}$ out to which dynamical tide resonances are detectable [$\Delta (\delta \phi_a) = |\delta \phi_a|$]  assuming a single merger event.}
\hspace*{-0.5cm}
\setlength{\tabcolsep}{0.15cm}
\begin{tabular}{cccccc}
\hline
 $f_a$ [Hz]  &$|\delta \phi_a|$ [rad]& Detector & $d_{\rm th}^{\rm hor}$ [Mpc] & SNR & $\Delta f_a$ [Hz] \\
\hline
%Adv. LIGO & 300 & 12 & $3.0e+4$  & 1.3  \\
	&	& aLIGO &  1.9 & 1.9e+3 & 2.4e+2 \\
450 & 0.01 & CE & 27 & 5.5e+3 & 2.1e+2  \\
	&	& ET-D & 17 & 2.5e+3 & 2.2e+2 \\
\hline
%Adv. LIGO & 300 & 12 & $3.0e+4$  & 1.3  \\
	&	& aLIGO &  0.19 & 1.9e+4 & 2.4e+2 \\
450 & 0.001 & CE & 2.7 & 5.5e+4 & 2.1e+2  \\
	&	& ET-D & 1.7 & 2.5e+4 & 2.2e+2 \\
\hline
	&	& aLIGO & 0.06 & 5.7e+4 & 2.5e+2 \\
750 & 0.001 & CE & 0.8 & 1.8e+5 & 2.3e+2 \\
	&	& ET-D & 0.5 & 8.1e+4 & 2.4e+2 \\
\hline
\end{tabular}
\end{center}
\end{table}

For the numerical results, we take $M=M'=1.4 M_\odot$ (using $M=M'=1.5 M_\odot$ or $1.6M_\odot$ in the Fisher matrix calculation changes the rms errors only by $\lesssim 10\%$).  Since the values of $t_c$ and $\phi_c$ do not affect the evaluation of $\Delta(\delta \phi)$ and $\Delta f_a$, we set them equal to zero. We show results for three sets of tidal parameters, one with $f_a=450$ Hz and $\delta \phi_a=0.01\textrm{ rad}$, corresponding to the lowest order muonic g-mode in the NS/HS model, one with $f_a=450\text{ Hz}$ and $\delta \phi_a=0.001\text{ rad}$ for the first hyperonic mode in the $1.5 M_\odot$ HS model, and one with $f_a=750 \text{ Hz}$ and $\delta \phi_a=0.001 \text{ rad}$ for the first hyperonic mode in $1.6 M_\odot$ HS. Higher order g-modes resonate at lower frequencies and have smaller $\Delta(\delta \phi_a)$ because more SNR is accumulated after their resonances [eq. (\ref{eq:dh_dphi_a})]. However, because of the steep falloff of $\delta \phi_a$ with decreasing frequency [increasing $n_a$; see eq. (\ref{eq:Qvsn})], it is much more difficult to detect their phase shifts. 

For the detector noise, we consider the noise curves of Advanced LIGO (aLIGO) at design sensitivity \citep{Aasi:15}, and the noise curves of proposed third generation detectors including the Cosmic Explorer (CE; \citealt{Evans:16}) and the Einstein Telescope (specifically, ET-D; \citealt{Hild:11}).  For simplicity we consider here only the detectability of a single detector instead of a network of detectors \citep{Schutz:11}. We also ignore the tidal phase shift due to the companion NS/HS which should increase $\delta \phi_a$ by a factor of two if $M=M'$. 

Our analysis neglects systematic uncertainties due to calibration errors in the instruments. The current calibration uncertainty of  aLIGO is somewhat larger than the phase shift due to dynamic tides (the phase uncertainty is $\simeq 0.03\textrm{ rad}$ at 450 Hz; \citealt{Vitale:12, LSCcal:16}). Regardless, we show that for aLIGO the measurement errors dominate ($\Delta(\delta \phi_a) \ga 0.1\textrm{ rad}$ even with event stacking) and preclude detecting the dynamical tide with aLIGO. As for the third generation detectors, currently there is no published estimate of their expected calibration performance. We therefore ignore this effect in our study and work under the assumption that it will be at least a factor of $\sim 10$ better than aLIGO's.

\subsubsection{Single events}

In Table \ref{tab:detectability} we show the threshold horizon distance $d_{\rm th}^{\rm hor}$ out to which different GW detectors can measure tidal resonances assuming a single merger event. Here $d_{\rm th}^{\rm hor}$ is defined to be the distance at which an event has $\Delta (\delta \phi_a) = \delta \phi_a$ assuming an optimal antenna response $\mathcal{K}=1$.  We also give the events corresponding SNR and $\Delta f_a$. For the dominating muonic mode, aLIGO can detect such a feature only for an event happening within 1.9 Mpc (i.e., within the Local Group). With the third generation detectors, this horizon distance can be pushed out to $\approx 20-30\textrm{ Mpc}$, and thus could include the Virgo Cluster ($16 \text{ Mpc}$). If we  account for the random source orientation (by dividing $d_{\rm th}^{\rm hor}$  by 2.3) and assume the `most-likely' event rate of $\mathcal{R}=10^{3}  \textrm{ Gpc$^{-3}$ yr$^{-1}$}$ (\citealt{Abadie:10, BNSrate:16}), we find that such an event should happen only once every $\approx 150\textrm{ yr}$ at  CE sensitivity (here we ignore that the Universe is far from being homogeneous in such a local range and simply assume that the sources are uniformly distributed). 

In terms of SNR, the detection of g-modes from a single event typically requires an $\text{SNR} \gtrsim 2000$, with the exact number depending on the detailed shape of the sensitivity curve. As a comparison, for a typical event with $\text{SNR}=12$, with aLIGO we can only measure $\Delta (\delta \phi_a) = 1.9 \textrm{ rad}$, which is more than two orders of magnitude larger than $\delta \phi_a$, and $\Delta f_a=4.7\times10^{4} \textrm{ Hz}$, which is greater than the entire detector bandwidth. 

As for the hyperonic mode, its small phase shift makes its detection possible only from the extremely loud events with SNR $\gtrsim 10^4$. Even with CE, the most sensitive detector we consider, the horizon distance can only reach 2.7 (0.8) Mpc for the $1.5 M_\odot$ ($1.6 M_\odot$) HS model. Therefore, unless there is an extremely rare nearby event, the phase shift due to a hyperonic mode is unlikely to be detected from  a single event.

\subsubsection{Multiple stacked events}

In Table \ref{tab:detectability_stack}, we give the rms errors $\langle \Delta(\delta \phi_a)\rangle$ and $\langle\Delta(f_a)\rangle$ found by stacking multiple events. We assume a total observation duration $P_{\rm obs}=5$ years, an SNR cutoff of 40 to determine $d_{\rm max}^{\rm eff}$, and an event rate of $\mathcal{R}=10^{3}  \textrm{ Gpc$^{-3}$ yr$^{-1}$}$\citep{Abadie:10, BNSrate:16}.  Under these assumptions, the number of expected events is 1, $9\times10^4$, and $2\times 10^3$ for aLIGO, CE, and ET-D, respectively (CE has the largest $d_{\rm max}^{\rm eff}$, with a cosmological redshift $z=0.3$).

We find that aLIGO can only measure phase shifts to an accuracy of $\langle\Delta (\delta \phi_a)\rangle \gtrsim 0.2 \textrm{ rad}$.  Since this is at least an order of magnitude larger than the phase shifts induced by resonant mode excitation, the dynamical tide is unlikely to be detectable with aLIGO even with event stacking. 

By contrast,  with CE the phase shifts can be measured to an accuracy of $\langle\Delta (\delta \phi_a)\rangle \approx 3\times 10^{-3} \textrm{ rad}$.  CE should therefore be able to detect the phase shift due to the muonic modes ($\delta \phi_a \approx 10^{-2}\textrm{ rad}$); they might only be marginally detectable with a single ET-D alone, however.  
CE has rms errors that are $\approx 100$ times smaller than aLIGO because its  $\Delta \theta_{\rm ref}$ is 15 times smaller (it accumulates 15 times more SNR post-resonance than aLIGO for the same source) and, for a given SNR cutoff,  its $d_{\rm max}^{\rm eff}$ is $\approx 40$ times larger [cf. eq. (\ref{eq:stack_events})]. 
%it is $\approx 20$ times more sensitive  and can thus detect $\approx 20^3=10^4$ more events in the same observing time.
 
Detecting the phase shift due to the hyperonic modes ($\delta \phi_a \approx 10^{-3}\textrm{ rad}$) will, however, be difficult even with CE event stacking given that $\langle\Delta (\delta \phi_a)\rangle \gtrsim \delta \phi_a $.  Moreover, $\mathcal{R}$ is predicted to be smaller for higher-mass NSs \citep{Kiziltan:13}, i.e., those containing hyperon cores, and therefore there will be fewer such events to stack.  It may also be difficult to distinguish the phase shifts of the hyperonic modes from those of the muonic modes, especially if the dominant (i.e., lowest order) muonic and hyperonic modes have similar frequencies, as is the case in our $1.5 M_\odot$ HS model.

%First of all, it requires a calibration accuracy that is at least 30 times better than the current level \citep{LSCcal:16}, which may be challenging. Besides, $\mathcal{R}$ is in fact predicted to be smaller for higher-mass NSs, i.e., those containing hyperon cores \citep{Kiziltan:13} and therefore less events to be stacked.  Furthermore, it may be difficult to distinguish the phase shifts of the hyperonic modes from those of the muonic modes, especially if the dominant (i.e., lowest order) muonic and hyperonic modes have similar frequencies, as is the case in our $1.5 M_\odot$ HS model.

%They might be marginally detectable if there are multiple third-generation detectors, with $P_{\rm obs} > 5\textrm{ years}$, or the event rate is at the high end of the predicted range ($\mathcal{R}=\textrm{few } \times 10^{3}  \textrm{ Gpc$^{-3}$ yr$^{-1}$}$; \citealt{BNSrate:16}). However, $\mathcal{R}$ is in fact predicted to be smaller for higher-mass NSs, i.e., those containing hyperon cores \citep{Kiziltan:13}.  Furthermore, it may be difficult to distinguish the phase shifts of the hyperonic modes from those of the muonic modes, especially if the dominant (i.e., lowest order) muonic and hyperonic modes have similar frequencies, as is the case in our $1.5 M_\odot$ HS model.

Our stacking calculation only accounts for the distance distribution of the sources but otherwise assumes all the events are identical.  It therefore neglects variation of the inspiral parameters, including the NS mass distribution.  Although this is a coarse approximation that should be relaxed in future studies, we do not expect it to significantly affect our estimates of $\langle \Delta   (\delta \phi_a)\rangle$.  From equation (\ref{eq:stack_events}), we see that $\langle \Delta  \theta \rangle \propto  \Delta \theta_{\rm ref}$, where $\Delta \theta_{\rm ref}$ is the reference value that is intended to be  representative of all the events.  Considering the phase shift measurement ($\theta = \delta \phi_a$), we showed in Section \ref {sec:detectability_model} that  $\Delta (\delta \phi_a)$ is independent of $\delta \phi_a$ and instead mostly depends on $f_a$. From  Table \ref{tab:delPhi} we see that for the muonic modes, $f_a$ is nearly independent of $M$ (it only changes be $10\%$ in going from $1.4 M_\odot$ to $1.6 M_\odot$).  Therefore, for each detector there is a reliable reference value for $\Delta (\delta \phi_a)$ and our estimate of $\langle \Delta   (\delta \phi_a)\rangle$ should not be strongly affected by our neglect of the mass distribution of the events.  
  
%Note that the stacking calculation assumes all the events are identical. Allowing the masses to vary will change $f_a$ and $\delta \phi_a$ for each event. However, for the muonic mode which is potentially detectable with the next generation detectors, changing mass from $1.4 M_\odot$ to $1.6 M_\odot$ only changes $f_a$ by $10\%$ (cf. Table \ref{tab:delPhi}). While $\delta \phi_a$ can vary by a factor of 3, from eqs. (\ref{eq:dh_dphi_a}), (\ref{eq:dh_df_a}), as well as Table \ref{tab:detectability_stack}, how well we can constrain $\Delta (\delta \phi_a)$ is independent of $\delta \phi_a$, and $\Delta f_a \propto (\delta \phi_a)^{-1}$.  Moreover, the mass distribution of NS in binary systems is likely to be narrowly peaked with scattering of only $0.1 M_\odot$. Therefore even if we average over a realistic distribution of events, $\langle\delta (f_a) \rangle$ is at most worsen by a factor of three while $\langle\Delta (\delta \phi_a) \rangle$ is unlikely to be significantly affected. For the hyperonic modes though, a real source distribution may make their detection even more challenging. 

\begin{table}
\begin{center}
\caption{\label{tab:detectability_stack} Measurement errors found by stacking events for different detectors and values of $f_a$ and $\delta \phi_a$.} 
%We assume 5 years of observation, an SNR cutoff of 12, and $\mathcal{R}=10^{3} \textrm{ Gpc$^{-3}$ yr$^{-1}$}$, corresponding to expected number of events of 45, $3\times 10^6$, and $8 \times 10^4$, for aLIGO, CE, and ET-D, respectively. }
\hspace*{-0.5cm}
\setlength{\tabcolsep}{0.15cm}
\begin{tabular}{ccccc}
\hline
 $f_a[\text{ Hz}]$  & $|\delta \phi_a|\text{ [rad]}$  & Detector &  $\langle{\Delta(f_a)}\rangle\text{ [Hz]}$  & $\langle\Delta(\delta \phi_a)\rangle\text{ [rad]}$ \\
\hline
	&	& aLIGO &  5.8e+3 & 2.5e-1 \\
450 & 0.01 &  CE &  5.3e+1 & 2.6e-3 \\
	&	& ET-D & 1.7e+2 & 7.7e-3 \\
\hline
	&	& aLIGO &  5.8e+4 & 2.5e-1 \\
450 & 0.001 & CE &  5.3e+2 & 2.6e-3 \\
	&	& ET-D & 1.7e+3 & 7.7e-3 \\
\hline
	&	& aLIGO & 1.8e+5 & 7.4e-1 \\
750 & 0.001 & CE & 2.0e+3 & 8.6e-3 \\
	&	& ET-D & 5.9e+3 & 2.5e-2 \\
\hline
\end{tabular}
\end{center}
\end{table}

\section{CONCLUSIONS}
\label{sec:conclusions}

We studied the dynamical tide in coalescing NS binaries and investigated how the resonant excitation of g-modes might impact the GW signal.  In our previous work (YW17), we carried out the first study of dynamical tides in a superfluid NS. We showed that the dynamical tide, unlike the equilibrium tide, is directly sensitive to the composition and superfluid state of the core  and thus offers a unique probe of the NS interior.  Here, we extended the results of YW17 by allowing for hyperons in the NS core.  \cite{Dommes:16} showed that  hyperons modify the buoyancy profile in the star and give rise to a new type of g-mode.  We confirmed their results, and calculated the spectrum of  hyperonic g-modes in the inner core and muonic g-modes in the outer core for different NS models (with  varying hyperon core radii $R_\Lambda$). We found that the characteristic frequency of the hyperonic g-modes increases linearly with $R_\Lambda$ and that the hyperonic g-modes can have considerably higher frequencies than the muonic g-modes.  We also showed that the frequency and tidal coupling of the muonic modes is not particularly sensitive to the existence of hyperons in the core.

The resonant tidal excitation of the hyperonic and muonic g-modes remove energy from the orbit and induce phase shifts in the GW signal.  We found that the lowest order g-modes induce the largest phase shifts, with magnitudes $|\delta \phi_a|\sim 10^{-3}-10^{-2}\textrm{ rad}$.  The muonic g-modes, which are concentrated in the outer core where the tide is stronger, induce phase shifts that are a few times larger than that of the hyperonic g-modes.  

Using the Fisher matrix formalism, we estimated the detectability of the induced phase shifts both from single events and from stacked events.  We found that with the next generation GW detectors (CE and/or ET), a single, optimally oriented event within $\approx 20\textrm{ Mpc}$ (e.g., within the Virgo Cluster) should be loud enough to detect the phase shift due to the muonic modes.  The system would need to be about five to ten times closer (e.g., within the Local Group), in order to detect the smaller phase shifts associated with the hyperonic modes.  While such nearby events are rare, we found that by stacking multiple events, there is a reasonably good likelihood that next generation detectors can detect the phase shifts induced by the muonic modes. Measuring the frequency and phase shift of the muonic mode resonance can help constrain the stratification and superfluid state of the NS core.  The phase shift due to the hyperonic modes will likely be difficult to detect even with event stacking. %unless the merger rate is on the high end of the predicted range.  
 
By restricting our models to relatively low masses ($M \le 1.6 M_\odot$), we ensured that the only type of hyperon in the core were $\Lambda$ hyperons.  In the future, it might be interesting to also study higher mass models, which would include $\Xi^-$ and $\Xi^0$ hyperons. Their composition gradients presumably give rise to yet more types of g-modes in the core.  We also assumed that the $\Lambda$ hyperons were normal fluid in the inner core (consistent with the treatment in \cite{Dommes:16}).  Whether the hyperons will be superfluid is uncertain, but if they are then they might either have a modified g-mode spectrum compared to our model or they might not support g-modes at all (similar to the g-modes supported by the proton-to-neutron gradient, which vanish when the neutrons are superfluid; \citealt{Lee:95, Andersson:01, Prix:02}).  Finally, it would also be interesting to investigate how tidal coupling in a superfluid NS is modified by rotation and nonlinear instabilities, both of which have been shown to be potentially important in the normal fluid case (see, respectively, \citealt{Ho:99, Lai:06, Flanagan:07} and \citealt{Weinberg:13, Venumadhav:14, Weinberg:16, Essick:16}).

\section*{Acknowledgements}
The authors thank Reed Essick, Tanja Hinderer, and the referee for valuable comments. This work is supported in part by NASA ATP grant NNX14AB40G. HY is also supported in part by the National Science Foundation and the LIGO Laboratory. LIGO was constructed by the California Institute of Technology and Massachusetts Institute of Technology with funding from the National Science Foundation and operates under cooperative agreement PHY-0757058.

%%%%%%%%%%%%%%%%%%%%%%%%%%%%%%%%%%%%%%%%%%%%%%%%%%

%%%%%%%%%%%%%%%%%%%% REFERENCES %%%%%%%%%%%%%%%%%%

% The best way to enter references is to use BibTeX:

\bibliographystyle{mnras}
\bibliography{ref}{} % if your bibtex file is called example.bib

%%%%%%%%%%%%%%%%%%%%%%%%%%%%%%%%%%%%%%%%%%%%%%%%%%

%%%%%%%%%%%%%%%%% APPENDICES %%%%%%%%%%%%%%%%%%%%%

\appendix
\section{RATE OF DIRECT URCA PROCESS}
\label{appendix:urca}
We show here that the timescale for a resonantly oscillating fluid element to relax toward chemical equilibrium by the $\Lambda$-hyperon direct Urca process
\begin{equation}
\Lambda \to {\rm p} + \rm{L} + \bar{\nu}_{\rm L} ,\hspace{0.5cm}
{\rm p} + \rm{L}  \to \Lambda + \nu_{\rm L}, \hspace{0.1cm} \textrm{ where L}=\rm{e}, \mu
\label{eq:reaction}
\end{equation}
is much longer than its oscillation timescale. While the nucleonic direct Urca (eq. \ref{eq:reaction} but with $\textrm{n}$ in place of $\Lambda$) is also possible in the core \citep{Lattimer:91}, its rate is greatly suppressed because the neutrons are superfluid and the core temperature is well-below the critical temperature for neutron superfluidity (see, e.g., \citealt{Yakovlev:01}).  Previous studies have shown that other damping mechanisms are also slow compared to the oscillation timescales we consider \citep{Reisenegger:92, Reisenegger:94, Lai:94}. We therefore conclude that the composition of the fluid element is nearly frozen and that the modes can be treated as adiabatic to a good approximation. 

Our calculation is similar to one given in\citeauthor{Yakovlev:01} (2001; Section 3.5). We focus on the reaction with $\rm{L}=\rm{e}$; the reaction with $\rm{L}=\mu$ occurs at a similar rate \citep{Prakash:92}. Note that while our calculation should hold for an NS with np$\Lambda$e$\mu$ composition, as is the case for the models we consider, for more massive NSs or NSs with different equation of state that also host other hyperon species, non-Urca weak interactions like $\rm{n}+\Lambda \rightleftharpoons \rm{p}+\Sigma^{-}$ can have a much higher reaction rate  than direct Urca processes and might therefore modify the calculation \citep{Yakovlev:01}.
 
We first define the chemical equilibrium parameter 
$
\beta_\Lambda=\mu_\Lambda - \mu_{\rm p} - \mu_{\rm e}
$
and the deviation from equilibrium relative to the background temperature
\begin{equation}
\eta = \frac{\delta \beta_\Lambda}{kT},
\end{equation}
where $\delta \beta_\Lambda = \delta \beta_\Lambda(P,\mu_{\rm n},x_{\mu {\rm e}}, x_{\Lambda {\rm e}})$. Although we are interested in the dis-equilibrium of a perturbed fluid element, and thus the Lagrangian perturbation $\Delta \beta_\Lambda$, since the background is in chemical equilibrium $\beta_\Lambda=0$ and therefore to linear order $\delta \beta_\Lambda = \Delta \beta_\Lambda$. 
%Since we our interested in small perturbations within a cold NS, with the typical value of $\delta \beta_\Lambda/k \lesssim 2\times 10^{10} \text{K}$, we are in the regime that $\eta\gg 1$ and  $\delta \beta \ll \mu_\Lambda$.

The timescale to relax toward chemical equilibrium is
\begin{equation}
\tau_{\rm urca} \approx \frac{\eta}{\dot{\eta}}.
\end{equation}
To proceed, we estimate $\tau_{\rm urca}$ in an iterative manner. First,  we obtain a zeroth order solution by assuming the composition is frozen and thus $\Delta x_{\Lambda \rm e}=0$ (and also $\Delta x_{\mu \rm e}=0$). We then use this solution to evaluate $\eta$ and $\tau_{\rm urca}$. As long as the resulting reaction timescale is much longer than the oscillation timescale, our approach is self-consistent. This approach implies
\begin{equation}
\eta \approx  \frac{1}{kT}\left(\frac{\partial \beta_\Lambda}{\partial x_{\Lambda \rm e}}\right)
%_{P, \mu_{\rm n}, x_{\mu \rm e}}  
\delta x_{\Lambda \rm e}
=
 -\frac{b_a}{kT} \left(\frac{\partial \beta_\Lambda}{\partial x_{\Lambda \rm e}}\right)\frac{\diff x_{\Lambda \rm e}}{\diff r} \xi_{\rm c}^r,
\end{equation}
where $b_a$ is the amplitude of the resonantly driven mode.
The first equality follows because $P$ and $\mu_n$ adjust to the background values on very short timescales, and $x_{\mu \rm e}$ is nearly constant in the inner core (cf. Fig.   \ref{fig:compGrad}). The second equality follows because in our zeroth order solution $\Delta x_{\Lambda \rm e} = 0$. In order to calculate  $\xi_{\rm c}^r$ and $b_a$, we use the numerical solution for the $n_a^{(\Lambda1.6)}=1$ mode, which has a post-resonance  amplitude $b_{a}\simeq 1\times10^{-3}$. We find that $\delta \beta_\Lambda \sim 2\textrm{ MeV} \gg kT$ for the largest perturbations and therefore for a cold NS we are in the limit $\eta \gg 1$.

Due to the Urca process, the deviation from chemical equilibrium changes at a rate 
\begin{equation}
\dot{\eta} \approx  \frac{1}{kT}\left(\frac{\partial \beta_\Lambda}{\partial x_{\Lambda \rm e}}\right)\dot{x}_{\Lambda \rm e},
\end{equation}
with
\begin{align}
\dot{x}_{\Lambda \rm e} = \frac{n_e\dot{n}_\Lambda - n_\Lambda \dot{n}_{\rm e}}{n_{\rm e}^2}=\frac{(1+x_{\Lambda \rm e})}{n_e}\dot{n}_\Lambda
= -\frac{(1+x_{\Lambda \rm e})}{n_e}\delta \Gamma.
\end{align}
The net reaction rate per unit volume  
\begin{align} 
\delta \Gamma&= \Gamma(\Lambda \to {\rm p} + \rm{e}^{-} + \bar{\nu}_{\rm e}) - \Gamma({\rm p} + {e}^{-} \to \Lambda + \nu_{\rm e})
\nonumber \\
&=q \frac{\epsilon}{kT} H(\eta)\eta
\label{eq:reaction_rate}
\end{align}
 (eqs. (144)-(147) in \citealt{Yakovlev:01}). 
 Here $\epsilon$ is the equilibrium neutrino emissivity [eq. (124) and Table 4 in \citealt{Yakovlev:01}; we use the rest masses rather than the effective masses, which is sufficiently accurate here] and is given by
 \begin{equation}
\epsilon=1.6\times10^{26}\left(\frac{n_\textrm{e}}{0.16\textrm{ fm}^{-3}}\right)\left(\frac{m_\Lambda}{m_{\rm N}}\right)\left(\frac{T}{10^9 \textrm{ K}}\right)^6\textrm{ erg cm$^{-3}$ s$^{-1}$}. 
\label{eq:emissivity}
\end{equation}
Although we calculate $\epsilon$ based on the hyperonic direct Urca proceess, we use the nucleon direct Urca values (assuming normal fluid nucleons) for the numerical prefactor $q=0.158$ and for the function 
\begin{equation} 
H(\eta)=1+\frac{10 \eta^2}{17\pi^2} + \frac{\eta^4}{17\pi^4} \simeq \frac{\eta^4}{17\pi^4}.
\end{equation}
The hyperon process should yield very similar values for $q$ and $H(\eta)$ [cf. Section 3.3 and eqs. (112)-(117) in \citealt{Yakovlev:01}; the energy integral $I$ is the same for all direct Urca processes and the angular integrals $A$ are similar].
 
Since the threshold of the hyperon direct Urca process is expected to nearly coincide with the threshold for the creation of hyperons  \citep{Prakash:92}, we assume that equation (\ref{eq:reaction_rate}) applies for $r\le R_\Lambda$.  We thereby find
\begin{align}
\tau_{\rm urca}^{-1}
&\approx q\left(\frac{\partial \beta_\Lambda}{\partial x_{\Lambda \rm e}}\right)\left(\frac{1+x_{\Lambda \rm e}}{n_{\rm e}}\right) \frac{\epsilon}{(kT)^2} H(\eta) \nonumber\\ 
&\approx 1.5\times10^{-2} \left(\frac{\delta \beta_{\Lambda}}{1 \textrm{ MeV}}\right)^4 \textrm{ s}^{-1}.
\label{eq:rate1}
\end{align} 
Although both $x_{\Lambda \rm e}$ and $\left({\partial \beta_\Lambda}/{\partial x_{\Lambda \rm e}}\right)$ depend on density $\rho$, to obtain the numerical result on the second line we treated them as constant since $\rho$ varies slowly in the inner core.  Note that $\tau_{\rm urca}^{-1}$ is independent of temperature in the limit $\eta \gg 1$.

Since $\tau_{\rm urca}^{-1}$ is much smaller than the oscillation frequency of the g-modes, we conclude that we can safely neglect the hyperonic direct Urca process and adopt the frozen composition approximation.

\section{PHASE SHIFT}
\label{appendix:phase}

In this appendix we derive the phase $\Psi(f)$ of the frequency-domain waveform $\tilde{h}(f)\propto \exp\left[i\Psi(f)\right]$ due to the resonant excitation of a single mode with eigenfrequency $f_a$. 
Following \citeauthor{Flanagan:07} (2007; see also \citealt{Reisenegger:94}), the resonance induces a sudden perturbation $\delta \dot{f}$ to the rate of frequency evolution relative to the point particle model
\begin{equation}
\dot{f}(f)=\dot{f}_{\rm pp}(f) + \delta \dot{f}(f).
\end{equation}
The time taken by the binary to evolve to some frequency $f$ after the resonance is, to linear order in $\delta \dot{f}$,
\begin{equation}
\label{eq:tf}
t(f) = \int_0^f \frac{\diff f}{\dot{f}} \simeq \int_0^f \frac{1}{\dot{f}_{\rm pp}}\left(1-\frac{\delta\dot{f}}{\dot{f}_{\rm pp}}\right)\diff f = t_{\rm pp}(f) - \delta t_a,
\end{equation}
where $\delta t_a= -\delta \phi_a/2\pi f_a$ (see Section IV of \citealt{Flanagan:07}; note that their sign convention for $\delta \phi_a$ is opposite ours).  Here we ignore the bandwidth of the resonance and treat the perturbation $\delta \dot{f}$ proportional to a delta function at $f=f_a$ since the duration of the resonance is much shorter than the orbital decay time scale (their ratio $\simeq 0.1\times[(\mathcal{M}/1.2M_\odot)(f_a/500 \text{ Hz})]^{5/6}$; \citealt{Lai:94}).  Since $\delta\phi_a < 0$, the mode resonance speeds up the inspiral and it takes slightly less time to reach a given post-resonance $f$.  After the resonance, the phase
\begin{equation}
\label{eq:phitf}
\frac{\phi[t(f)]}{2\pi}\hspace{-0.05cm} =\hspace{-0.05cm}  \int_0^{t(f)}\hspace{-0.4cm}f(t')\diff t' \hspace{-0.05cm}\simeq \hspace{-0.05cm}\int_0^f \hspace{-0.05cm} \left[\frac{f'}{\dot{f}_{\rm pp}} - \frac{f' \delta\dot{f}}{\dot{f}_{\rm pp}^2}\right]\diff f'\hspace{-0.1cm}=\hspace{-0.1cm}\frac{\phi_{\rm pp}(f) + \delta \phi_a}{2\pi}.
\end{equation}
The phase of the frequency domain waveform post-resonance is therefore
\begin{equation}
\Psi(f) = 2\pi f t(f) - \phi[t(f)] - \frac{\pi}{4} = \Psi_{\rm pp}(f) - \left(1-\frac{f}{f_a}\right)\delta \phi_a,
\end{equation}
where the first equality comes from the stationary phase approximation and $\Psi_{\rm pp}= 2\pi f t_{\rm pp}(f)-\phi_{\rm pp}(f) - \pi/4$ is given by equation (\ref{eq:phi_pp}).  If we had chosen the parameters so that the perturbed and unperturbed waveforms coincide after the resonance rather than before the resonance then the $(1-f/f_a)$ factor would have the opposite sign. For the small $\delta \phi_a$ due to g-mode resonances, we find that both choices give nearly identical rms errors $\Delta(\delta \phi_a)$ and $\Delta f_a$ due to the covariance with $t_c$ and $\phi_c$.

%%%%%%%%%%%%%%%%%%%%%%%%%%%%%%%%%%%%%%%%%%%%%%%%%%

% Don't change these lines
\bsp	% typesetting comment
\label{lastpage}
\end{document}